\documentclass{article}

\usepackage{PRIMEarxiv}

\usepackage[utf8]{inputenc} 
\usepackage[T1]{fontenc}    

\usepackage{url}            
\usepackage{booktabs}       
\usepackage{amsfonts}       
\usepackage{nicefrac}       
\usepackage{microtype}      
\usepackage{lipsum}
\usepackage{fancyhdr}       
\usepackage{graphicx}       
\graphicspath{{media/}}     
\usepackage[colorlinks,urlcolor=blue,linkcolor=blue,citecolor=blue]{hyperref}

\title{Voluminous yet Vacuous? \\ Semantic Capital in an Age of Large Language Models
}

\author{
  Luca Nannini \\
  \texttt{lucanannini2019@gmail.com} \\
}

\begin{document}

\maketitle

\begin{abstract}
Large Language Models (LLMs) have emerged as transformative forces in the realm of natural language processing, wielding the power to generate human-like text. However, despite their potential for content creation, they carry the risk of eroding our Semantic Capital (SC) - the collective knowledge within our digital ecosystem - thereby posing diverse social epistemic challenges.
This paper explores the evolution, capabilities, and limitations of these models, while highlighting ethical concerns they raise.
The study contribution is two-fold: first, it is acknowledged that, withstanding the challenges of tracking and controlling LLM impacts, it is necessary to reconsider our interaction with these AI technologies and the narratives that form public perception of them.
It is argued that before achieving this goal, it is essential to confront a potential deontological tipping point in an increasing AI-driven infosphere. This goes beyond just adhering to AI ethical norms or regulations and requires understanding the spectrum of social epistemic risks LLMs might bring to our collective SC.
Secondly, building on Luciano Floridi's taxonomy for SC risks, those are mapped within the functionality and constraints of LLMs.
By this outlook, we aim to protect and enrich our SC while fostering a collaborative environment between humans and AI that augments human intelligence rather than replacing it.
\end{abstract}

\section{Introduction}

The fable of Funes the Memorious, conceived by Jorge Luis Borges, serves as a powerful metaphor for the era we live in. Funes, the character blessed—or rather, cursed—with perfect memory, found himself submerged in an ocean of unfiltered details. He was a prisoner of his own capacity, drowning in his universe of relentless particulars. The individual who once boasted the greatest memory lost his ability to discern the important from the trivial, transforming his mind into a "\textit{garbage heap}" of excessive detail. As Borges wrote, "\textit{To think is to forget differences, generalize, make abstractions. In the teeming world of Funes, there were only details, almost immediate in their presence}" \cite{borges1962funes}. In an echoing resonance to Funes' plight, our society now finds itself amidst a surge of information generation and consumption with uncharted challenges to our epistemic filters.

This era, marked by the infinite details of our rapidly expanding infosphere, mirrors Funes' predicament. In this context, the concept of \textbf{Semantic Capital} (SC), coined by Luciano Floridi, gains paramount importance. SC encapsulates the collective information resources—knowledge, skills, or competencies—that individuals or entities possess. These resources can be harnessed to create value within our interconnected global information ecosystem \cite{floridi2010information}. This construct actively shapes the infosphere, catalyzing communication, fostering innovation, and driving informed decision-making \cite{floridi2016semantic}.
As the realms of human cognition and artificial intelligence (AI) increasingly coalesce, their intersection is redefining the landscapes of collaboration, decision-making, and knowledge creation. In these emerging dynamics, the role of SC escalates in importance and complexity.
The collaborative environments entailing human and AI integration go beyond mere task execution; they embody intricate interactions that should be dependent on mutual beneficial augmentation, and not replacement or mimicry \cite{brynjolfsson2022promise, Vallor2022}.
Nonetheless, this surge in information, fueled in part by AI, has the potential to generate a cascade of cognitive and sociotechnical risks e.g., cognitive overload, misinformation, social polarization, and erosion of public trust. As we argue in this paper, we are possibly approaching a deontological 'tipping point'—an inflection where our moral obligation to promote open dissemination of AI information might conflict with our duty to prevent harm. Indeed, the relentless acceleration and proliferation of AI information might soon manifest their most detrimental and repressive effects \cite{han2017swarm, crawford2021atlas, bridle2018new, mcquillan2022resisting}.

This paper endeavors to delve into the role of SC within the sphere of human-AI interaction. We depart by defining its value to foster societal knowledge and trust within the context of the information ethics challenges that we currently face. We pay particular attention to generative AI systems, particularly Large Language Models (LLMs).
By contextualizing the debate happening over their capabilities and limitation, we move to address the broader range of ethical and deontological implications of LLMs. Central to this endeavor is a necessary reframing of AI narratives, with a mindful consideration of who benefits from these narratives and how they shape public perception.
In an era where our infosphere is populated with increasingly accessible AI-generated content, a critical reassessment of our relationship with open-source practices is paramount. By reflecting on the value of open-source models and regulations, we endorse governance practices to ensure they align with our ethical obligations and societal values. 
But before achieving that, we exhort to consider a deontological tipping point in our AI-driven infosphere. This approach entails moving beyond calls to adhere to AI ethical guidelines or regulations and conceiving a range of social epistemic risks that LLMs pose to our collective SC. Following Floridi's taxonomy for SC risk, our main contribution lies in mapping them within the capabilities and limitations of LLMs. By doing so,  we aim to bring a novel outlook to the LLM discussion while encouraging innovative strategies to reinforce our epistemic defenses. Our discourse seeks to guide us toward an equitable and sustainable infosphere, where innovation flourishes without compromising societal values and individual well-being.


\section{Appraising Semantic Capital}

Our exploration of the crucial role of SC in human-AI collaboration begins with Luciano Floridi's philosophy of information. Floridi's seminal work, birthed from the metamorphosis of the information age, places information at the core of our world understanding \cite{floridi2011philosophy}. 
The \textit{infosphere}, Floridi proposes, is an immersive information environment housing all informational entities—humans, artificial agents, and other organisms \cite{floridi2011philosophy}. In this sphere, constant streams and exchanges of information form a complex interaction network, shaping our reality perception and directing our actions.
Within this infosphere nests SC—value derived from meaningful information. It transcends mere data accumulation, presenting as well-formed, meaningful data that bolsters one's power to create meaning—\textit{semanticise}. An individual's, group's, or society's SC stock, demonstrated in various forms like knowledge repositories, skills, shared societal norms, cultural narratives, etc., is employed and invested in information creation, understanding, and dissemination. This process fuels essential life aspects like communication, decision-making, learning, problem-solving, and others.
SC's value is intrinsically linked to its ability to enrich our understanding, navigation, and shaping of our realities. As such, managing and curating SC is vital in our increasingly information-dense society. The risks associated with SC— (a) loss, (b) unproductive, (c) underuse, or (d) misuse or or (e) depreciation due to truth erosion — are defined by Floridi as "\textit{the potential of loss of part or all of the value of some content that can no longer enhance someone’s power to semanticise something}".
\cite{floridi2011philosophy, floridi2016semantic}.

The digital technology era has brought forth new SC dimensions. Data abundance and computational power have created pathways for enhancing and expanding our SC. AI and other digital technologies facilitate SC management and curation, aiding its effective and efficient usage and enrichment. If our world understanding is based on relationships between informational entities, and not just their intrinsic properties \cite{floridi2016handbook}, then these technologies give rise to new SC forms that significantly impact our semanticising processes and, ultimately, shape our identities and realities\footnote{SC can be differentiated from related concepts like 'intellectual capital' and 'cultural capital'. While SC focuses on knowledge, skills, and resources used for communication and comprehension \cite{floridi2016semantic}, intellectual capital pertains to an organization's sum of knowledge and skills that provide a competitive edge \cite{stewart1997intellectual, firer2003intellectual}. Cultural capital, however, refers to cultural resources like education and norms that influence individual behavior and societal opportunities \cite{bourdieu1986forms}.}.
Why, then, is it essential to highlight these concepts? Their role in shaping human-AI collaboration is central. SC provides a crucial lens through which we understand, navigate, and shape the evolving landscape of human-AI interaction in the generative AI era.

\section{Development of LLMs and the Debate Over Language "Understanding"}


In the compelling narrative of natural language processing (NLP), we've borne witness to a series of remarkable advancements over the past decade, with Large Language Models (LLMs) and other AI generative systems claiming center stage \cite{bender2021dangers}. Commencing with the invention of Long Short-Term Memory (LSTM) networks in 1997 \cite{hochreiter1997long}, the journey has led us to the present-day marvels of AI, such as GPT-4 \cite{radford2023gpt4}. These developments have profound implications for SC, raising pressing questions about the deontology of knowledge and information resources within the infosphere.

A brief historical overview of NLP highlights the rapid progress and increasing complexity of these models. From LSTM to Word2Vec \cite{mikolov2013efficient}, from Sequence-to-Sequence models \cite{sutskever2014sequence} to the transformative attention mechanism \cite{bahdanau2014neural}, and ultimately to the groundbreaking Transformer architecture \cite{vaswani2017attention} and subsequent birth of BERT \cite{devlin2018bert}, each evolution has refined the capacity to process and generate text, thereby influencing the constitution and use of SC. 
The 'philosophical' foundations of these NLP applications, especially for Word2Vec, relied on concepts of Distributional Semantics \cite{firth1957synopsis, BrunilaL22}, paired with the untapped benefits and dangers of "\textit{Big Data}" to mirror a presumptive realistic image of textual knowledge gathered from online repositories and communities \cite{mayer2013big, Gandomi2015, ZookBarocas2015}. Against such shallow reflection, scholars addressed concerns related to biases of this knowledge available online or within any other databases with spurious, impartial, or unguarded data entries \cite{boydcrawford2012, calude2017deluge, mittelstadt2016ethics, navigli2023llms}. This raised challenges for these models, such as primarily avoid to display semantically incomplete or nonfactual information\footnote{the so-called "hallucinations" \cite{
ziweihallucination2023} in Natural Language Generation (NLG).}.
The advent of LLMs such as OpenAI's GPTs, and their deployment in various applications, represent the contemporary zenith of this technological trajectory \cite{radford2018improving}. Nevertheless, the rapid proliferation of these models has sparked a lively debate among researchers and scholars concerning their true capabilities and implications.

A crucial question raised in this debate is whether LLMs genuinely understand the information they process, or if they are mere "\textit{stochastic parrots}," as posited by AI researchers Emily Bender, Timnit Gebru, Angelina McMillan, and (under an alias) Margaret Mitchell \cite{bender2021dangers}. The paper offered a continuation of a critical inquiry toward their natural language understanding, as previously expressed in 2020 by Bender \cite{Benderclimbing2020}. They argued that these models, despite their seemingly human-like text generation abilities, merely mimic patterns without comprehending the underlying meaning, potentially leading to the dilution of SC shuffling human knowledge in a convincing manner. 
Foremost, their concerns were grounded around the biases embedded in the training data, the substantial environmental footprint of training such language models, and the concentration of power in a few tech giants controlling them. In echoing these concerns, Melanie Mitchell highlighted in December 2022 the limitations of LLMs in truly "\textit{understanding}" the world and their reliance on superficial patterns in the data \cite{melaniemitchell2022}.

Yet, it needs to be recognized how LLMs are powerful tools that generate human-like narratives: their underlying architecture and scalability allow them to manipulate and operate inferences over the external world representations. But such abilities are generally hard to forecast, as well as to handle and interpret, by their designers. The so-called "\textit{emergent abilities}," which become more evident as the scale of the models increases, refer to the unforeseen and unplanned behavior that LLMs display, which often defy easy understanding or control by the developers themselves \cite{bowman2023eight}. Such abilities can result in outputs that are surprisingly insightful or disturbingly off-mark, underscoring the unpredictability and potential risks of deploying LLMs in real-world contexts \cite{brown2020language, lietal2021implicit, schaeffer2023emergent, micelibarone2023larger}. This challenge intensifies when we consider the increasing number of studies being released for their application in practical scenarios to assist various human tasks \cite{wang2023describeexplain, xie2023llms}.

This translates to the fact that despite the property to handle a certain degree of semantic information \cite{lietal2021implicit} to produce coherent textual information, LLMs cannot be universally trusted as epistemic agents capable to handle pragmatic constraints of human communication. The reason for this lies in their architectures per sé, but also within potential \textit{Eliza-effect} \cite{melaniemitchell2022}, e.g. how the user linguistically frames their prompts based on their intention and competencies \cite{perez2022}. This entails that the presumptive factuality of these model outputs needs then to be compared against their stochastic nature, heavily influenced by the design \cite{turpin2023language, schaeffer2023emergent} and also the interaction \cite{zhang2023language} of the users with the prompts fed. Despite growing efforts in providing additional heuristic bases to downplay unpredictable behavior, such as with \textit{chain-of-thought}, \textit{constitutional AI} or \textit{red-teaming} \cite{bowman2023eight, wei2023chainofthought, wang2022}, a crucial question stands with the reliability and factuality of LLMs: can we equate the performances of LLMs with human understanding and knowledge? Recognizing the differences, the academic community is reevaluating how to benchmark these models' performance. This calls for more critical assessment measures that better reflect the nature and capabilities of LLMs, especially in terms of interpretability and predictability \cite{bowman2022dangers, tedeschi2023whats, turpin2023language, navigli2023llms}.

\section{Fear Sells Well? On Ethical and Deontological Implications of LLMs}

The debate over LLMs' abilities underscores the complex implications of AI generative systems for SC. Indeed, it comes as no surprise how LLMs, by their design and capabilities, can profoundly influence the infosphere landscape: they operate as powerful amplifiers and conduits of information, capable of synthesizing and generating vast amounts of text that are, in many instances, indistinguishable from human-written content.

Reasoning about their potential benefits, LLMs can democratize access to information by breaking down barriers to user understanding e.g., paraphrasing, summarizing, or translating text into different languages. By making information more accessible and interpretable, these models can enhance the inclusivity and utility of SC. Secondly, LLMs can also contribute to the expansion of SC by facilitating the creation of new content. Authors can use these tools to overcome writer's block, generate creative ideas, or automate routine writing tasks. In academic and professional settings, LLMs can help to compile emails, draft reports, write code, or even create poetry and prose, thereby enriching the diversity and volume of SC.
%
Such positive scenarios must be counterbalanced with a sober recognition of the potential costs and risks that these models pose to SC. Among others, the risks associated with LLMs extend beyond semantic information handling, touching upon socio-economic, political, and ethical domains, encompassing bias propagation, labor market disruptions, power centralization, misinformation campaigns, cyber threats, intellectual property issues, and unforeseen harmful uses \cite{weidinger2022risks}.

At the core, LLMs can potentially spawn a proliferation of information a-like content, increasingly blurring the line with factual information. This proliferation risks diluting the quality of SC, contributing to an infosphere that is voluminous yet vacuous\footnote{Not only related to textual abilities, but so far public opinion was surprised by the dissemination of hyperrealist portraits of public personas made through generative AI tools, e.g. Pope Francis [\href{https://www.nytimes.com/2023/04/08/technology/ai-photos-pope-francis.html}{TheNYT2023a}] wearing fashion coats or Donald Trump getting arrested [\href{https://www.forbes.com/sites/mattnovak/2023/03/19/viral-images-of-donald-trump-getting-arrested-are-totally-fake/}{Forbes2023}] getting arrested. Afterward, part of the public was enraged to see how a professional photographer, Boris Eldagsen, could even win an international award with an AI-produced image [\href{https://www.theguardian.com/technology/2023/apr/17/photographer-admits-prize-winning-image-was-ai-generated}{TheGuardian2023a}], or, even worse, famous painters such as Edward Hopper being displayed in Google's search engine alongside AI-generated imitations of their works [\href{https://futurism.com/top-google-result-edward-hopper-ai-generated-fake}{Futurism2023}].}.
Alongside these concerns, the susceptibility of LLMs to the propagation of false information, as explored by Bian et al., adds another layer of complexity to the debate \cite{bian2023drop}. Their study claimed how false information tends to spread and contaminate related memories in LLMs via a semantic diffusion process. Also, these models might be subject to authority bias, often accepting false information presented in a more trustworthy style such as news or research papers. 
On this line, if LLMs are easily perturbable given prompt and information sources provided, they might be deployed at scale to scuffle or crowd out minor or dissenting public voices\footnote{On this note, a debate should be held on how appropriate is to deploy generative AI to represent social distress and identities, such as public manifestations [\href{https://www.theguardian.com/world/2023/may/02/amnesty-international-ai-generated-images-criticism}{TheGuardian2023b}, \href{https://twitter.com/DeanSamed/status/1658833605882265602}{Twitter2023}], or companies recoursing to generative AI tools claiming to promote "\textit{diversity}" through fake fashion models advertising [\href{https://www.independent.co.uk/life-style/fashion/levis-ai-models-diversity-backlash-b2310280.html}{TheIndependent2023}], while in reality displaying a subtle operation of ethics-washing - failing to hire and remunerate underrepresented individuals while still leveraging their image at no costs.}.
Within these considerations, we should now approach the paper from Bender et al. \cite{bender2021dangers} as a starting point of a wider debate, encompassing not only capabilities of LLMs, but rather the governance implication and social communities impacted, ultimately pertaining to the value of our shared SC \cite{boyd2018}. 
Put in simple terms, their discourse shall be not considered merely a matter of academic disquisition over semantic handling of human language, but rather a pointed attempt to scrutinize how these generative tools are 
associated to a narrative about AI that serves those who possess the means and resources to develop and capitalize economic value and competitive advantage - not only directly from such models, but also around them \cite{stewart1997intellectual}.
Through this lens, it can be discerned two key interpretive perspectives in this debate. 
The first, immediate perspective mesmerizes the public by proclaiming these LLMs as "\textit{sparks of Artificial General Intelligence }(AGI)," \cite{bubeck2023sparks}, implying that these models display initial prototypes of human-a-like cognitive intelligence\footnote
{Yet, the research from Bubeck et al. is released [as for May 2023] without peer-review by a team of Microsoft and OpenAI's researchers, using foremost controversial definitions of human intelligence as a comparison [\href{https://www.nytimes.com/2023/05/16/technology/microsoft-ai-human-reasoning.html}{TheNYT2023b}]. Related to the AGI narrative, Giada Pistilli, main ethicist of HuggingFace and contributor of the LLM BLOOM \cite{workshop2023bloom}, claimed in May 2022 to not engage herself to speak any more of AGI in a fortunate Twitter thread [\href{https://twitter.com/GiadaPistilli/status/1530136739959951361}{Twitter2022}].This is because the framing of that public debate was proved only detrimental to the real harms of LLMs, cautioning an in-depth analysis of the issue in a research study published the same month \cite{Pistilli2022}. This position resonates with an increasing number of scholars being cautious to adopt or even engage in using these terms in the public discourse; similarly - as also for the current paper - concerns over unnecessary anthropomorphisms \cite{shanahan2023talking} within LLMs are now being raised while deploying terms pertaining to human cognition, such as "\textit{hallucination}", to address nonfactual information provided by LLMs [\href{https://www.theguardian.com/commentisfree/2023/may/08/ai-machines-hallucinating-naomi-klein}{TheGuardian2023c}]. In this perspective, some work is finally moving towards making explicit design choices to prevent anthropomorphism for conversational systems \cite{abercrombie2023mirages}.}.
Such a view captivates public imagination and fuels, at best, a techno-optimistic narrative, while at worst, technological determinism, having public opinion feel humanity as doomed by the advent of some unavoidable and imminent superior AI \cite{boyd2018}.
%
The second perspective, however, is way more sobering and less sensationalist, unpacking a far more structural and intricate argument concerning the ecology of AI development, commercialization, and the possession of SC in the form of know-how for gathering and maintaining increasingly sophisticated data and AI models \cite{crawford2021atlas}.
When the conversation revolves solely around the inherent risks in the models, it inadvertently diminishes the role of their developers. As Bender et al. resonated, their research served as a warning bell, cautioning against a development trajectory of AI solutions promising extraordinary capabilities without due scrutiny \cite{bender2021dangers, morozov2013save}.

The core issue resides in the polarization of a debate where, on one hand, one faction predominantly comprises stakeholders - such as proprietaries of AI solutions - might derive benefits from gauging public attention over these models. Their strategic maneuvers, despite genuine fears over downsides of their products, might also be geared towards maintaining the undivided attention of the global audience, intending to foster an environment conducive to the promotion and consumption of their AI-based creations. Concurrently, another group emerges posing stark opposition by unearthing the contentious aspects of such models. This group, yet widely heterogeneous, contends that these AI solutions are not inherently superior or advantageous, and instead, might cause more harm than good due to their pronounced socio-technical ramifications and the plausible monopoly \cite{mcquillan2022resisting} they can create in the AI innovation landscape
\footnote{
    Interestingly, the fervor and dynamism of this debate have garnered widespread attention. With the current momentum, an increasing number of scholars and civil rights associations are echoing the apprehensions about the potential LLMs can inflict, taking actions such as open letters to regulate LLMs.
    Of this group, a segment of the public is lending credence to the "\textit{longtermism}" outlook—holding onto the belief that AI might be a blessing for all humanity in the future, only if it is perceived as an existential threat today [\href{https://www.forbes.com/sites/lanceeliot/2022/10/25/ai-ethics-and-ai-law-wrestling-with-ai-longtermism-versus-the-here-and-now/?sh=3a4b515e1c0f}{Forbes2022}],[\href{https://www.bloomberg.com/opinion/articles/2023-05-19/ai-longtermism-alarmists-are-dragging-us-all-down-existential-rabbit-hole}{Bloomberg2023}]. This viewpoint, however, does not advocate for immediate and tangible action against present structural issues, such as the exploitation of underrepresented communities involved in annotating and moderating LLMs.
    In response to these systemic issues, the affected communities have begun showcasing innovative grassroots initiatives. Karen Hao's investigation into AI colonialism [\href{https://www.technologyreview.com/supertopic/ai-colonialism-supertopic/}{MIT, TechRev2022}] and the protests staged by African AI workers to unionize in Nairobi illuminate these ongoing efforts [\href{https://time.com/6275995/chatgpt-facebook-african-workers-union/}{Time2023}].
    Meanwhile, it is noteworthy that AI pioneers, like Geoffrey Hinton, have been vocal about the necessity for increased regulations but have not explicitly extended support to these communities or other concerned academics, such as Bender, Gebru, and Mitchell [\href{https://www.theguardian.com/technology/2023/may/02/geoffrey-hinton-godfather-of-ai-quits-google-warns-dangers-of-machine-learning}{TheGuardian2022}]. Similarly, owners of AI technologies, like OpenAI's CEO Samuel Altman, have sought regulatory measures before the US Senate [\href{https://www.nytimes.com/2023/05/16/technology/openai-altman-artificial-intelligence-regulation.html}{TheNYT2023}], while other industry leaders, such as Microsoft Chief Economist Michael Schwarz [\href{https://arstechnica.com/tech-policy/2023/05/meaningful-harm-from-ai-necessary-before-regulation-says-microsoft-exec/}{ArsTechnica2023}] and former Google CEO Eric Schmidt [\href{https://fortune.com/2023/05/15/former-google-ceo-eric-schmidt-tells-government-to-leave-regulation-of-ai-to-big-tech-openai-chatgpt-bardai-midjourney/}{Fortune2023}], have either invited caution over the perceived risks of generative AI until incidents of "\textit{meaningful harm}" occur or advocated for self-regulation in the industry while criticizing governments for their alleged lack of expertise to regulate technology effectively. The narrative spun by these AI proprietors oscillates between demanding no regulation and advocating for a different regulation. Such a seemingly contradictory stance might be interpreted as a strategic maneuver to hold investor attention captive while cleverly deflecting competitive threats in the AI arena [\href{https://www.businessinsider.com/ai-technology-chatgpt-silicon-valley-save-business-stock-market-jobs-2023-5}{Insider2023}].
}.
Indeed, the year 2023 witnessed an unprecedented surge in the release of LLMs applications to the public by large corporations.
These developments were characterized by increasingly shortened time-to-market durations, intensifying the potential risks and implications of these systems\footnote{The rush to launch these applications often eclipsed necessary precautions, resulting in technology releases without sufficient safeguards. This haste raises concerns about corporate decision-making and leaves the public exposed to unanticipated AI-related risks, such as LLMs chat-bots harassing or recommending users to self-harm or indulge minors into socially irresponsible behaviors
[\href{https://www.businessinsider.com/widow-accuses-ai-chatbot-reason-husband-kill-himself-2023-4}{Insider2023b}, \href{https://www.washingtonpost.com/technology/2023/03/14/snapchat-myai/}{WashinghtonPost2023}, \href{https://time.com/6256529/bing-openai-chatgpt-danger-alignment/}{Time2023}].
}.
This speed, while demonstrating their technological capabilities, also exposed gaps in their ethical governance. Despite their "\textit{demo}" status, instances of these LLMs causing harm or harassment to users highlighted the need for careful deployment strategies and comprehensive product testing and feedback, as well as structural inquiry over the influence exerted over the AI development agenda by proprietary solutions.

Such unforeseen detrimental consequences serve as stark reminders of the need to couple AI development with comprehensive evaluation processes that prioritize societal well-being over speed and profit. Navigating this debate, one must remain cognizant of the intricate dynamics at play and question who ultimately benefits from these narratives. This to ensure that the discourse around AI and its impact on our collective SC remains grounded in empirical realities and is sensitive to the broader socio-economic implications.

\section{Open Source and Regulation for LLMs}

Let's momentarily pause and look beyond the current maelstrom of the ongoing debate on LLMs. Taking a step back, we find ourselves in the birth of the internet era, deeply influenced by the late 20th century's internet narratives. This was a time ripe with the promise of an information revolution \cite{croff2006}. The birth of the open-source paradigm during this time served as a catalyst to this revolution. 
By providing a universal platform accessible to anyone with an internet connection, it was an embodiment of the democratic ethos of these emerging digital utopias. 
The open-source movement, anchored in collaboration, transparency, and accessibility, has spurred an incredible acceleration in technological evolution \cite{dibona1999open}. This movement's transformative impact is especially palpable in the AI field, cultivating a fertile ecosystem ripe for progress and innovation. Emerging in this backdrop, LLMs owe much of their rapid development to open-source AI frameworks like TensorFlow and PyTorch as well as the Transformer architecture \cite{NEURIPS2019_9015, abadi2016tensorflow, vaswani2017attention}. Such open-source tools have made it feasible for researchers, developers, and organizations across the globe to access, modify, and contribute to a shared body of knowledge and codebase. This democratization of AI technologies, however, is a double-edged sword; while it empowers innovation and progress, it simultaneously amplifies challenges related to misuse, ethical implications, and regulatory requirements.
The diffusion of generative AI technologies, such as LLMs, via open-source platforms, accentuates the dual-use risk. LLMs can be applied for both beneficial and harmful purposes. Still cognizant of their risks, once an AI model is made openly available, specularly becomes harder to track, contain, or retract, given the scale, speed, and accessibility facilitated by open-source platforms. If instead an LLM is proprietary, such as GPT-4 \cite{radford2023gpt4}, being undisclosed to the public, then risks might arise in not being able to reprove its design phase and data provenance, as well as oversight its deployment.

From this, it comes as no surprise that regulating generative AI technologies is a formidable challenge. The pace at which AI evolves is often unmatched by the rate at which traditional regulatory frameworks adapt\footnote{A lively example of this challenge can be found in the EU commissions efforts back in April 2023 to make amendments targeting generative AI, ahead of final parliamentary votation on May 11th with the EU AI Act draft [\href{https://www.europarl.europa.eu/news/en/press-room/20230505IPR84904/ai-act-a-step-closer-to-the-first-rules-on-artificial-intelligence}{EuroparlPress2023}, \href{https://www.euractiv.com/section/artificial-intelligence/news/ai-act-moves-ahead-in-eu-parliament-with-key-committee-vote/}{Euractiv2023}].}. Crafting effective regulations requires a delicate balancing act: on one side, for disclosed models, it entails to manage the risks of misuse while preserving the democratic ethos of open-source, without stifling innovation; on the other, for proprietary models, it entails preserving marketing advantages while still allowing impart auditing measures to reprove model compliance and benevolence within regulatory standards as well as societal values.
One potential pathway forward involves revisiting our relationship with open-source practices in the context of LLMs. This rethinking requires a comprehensive, integrative approach that respects the principles of open-source while recognizing and addressing the risks posed by AI technologies.
Strategies could include more accountable deployers' practices, having them bear a greater responsibility for their creations, and revised legal frameworks that adapt to the specific challenges of LLMs. In terms of soft-power, this could be complemented by industry-wide certifications and licensing\footnote{Within license, a leading example is RAILS. The BigScience project, an open collaborative initiative, introduces a Responsible AI License (RAIL) for the usage of their LLMs to balance accessibility and risk mitigation. It reflects a community-led approach to restrict potential LLM harms, such also concerns about their societal and environmental impacts [\href{https://bigscience.huggingface.co/blog/the-bigscience-rail-license}{BigScience-RAILS2023}].} to enhance accountability over the design and development of those AI systems.
In terms of hard-power, instead, AI governance measures should attain from clear legislative guardrails, such as regulatory sandboxes, risk assessments, and auditing practicing encompassing the development and deployment of LLMs. Within this scope, the current major regulatory effort in the global landscape is now being lead by the European Union (EU), yet not being exempted from potential legislative weaknesses that might not always efficiently mitigate LLMs risks\footnote{In particular, the current amendment draft of the EU AI Act voted on May 2023 introduced definition and provisions targeting LLMs, intended as \textit{foundation models} \cite{draftcompromise}. At the current stage of draft, Art.28b(4), although partially beneficial with its transparency obligations, is criticized for its lack of duties imposed on online AI content generators, necessary for curbing misinformation. Yet, the Act is not yet enforced, and will likely have to interplay, within the EU regulatory ecosystem, with other regulations being discussed or already enacted. For an in-depth overview of these legislative implications, also outstanding the EU ground, refer to the working paper of \cite{hacker2023}.}.   

\section{The Deontological Tipping Point: Navigating the Information Surge}


Yet, calling for ethical virtuosism and regulations might not be enough to shelter us our epistemic filters in this unprecedented storm of AI-generated information. While this surge of information has democratized access to knowledge and fueled progress in myriad fields, it also has the potential to create a state of social epistemic bewilderment.
It is against this backdrop that it can be argued that we have reached a \textit{deontological tipping point}—an inflection where the relentless acceleration and proliferation of information culminate in the epistemic condition to scale up its detrimental effects. The concept of a deontological tipping point suggested is constituted by a juncture where our moral obligation to assist to the open dissemination of certain AI narratives and solutions may come into conflict with our duty to prevent harm.
Within the context of AI, and particularly in relation to LLMs, this tipping point is precipitated by the realization that unfettered access to information and open-source practices, while fostering innovation, can also amplify risks given how scalable and accessible these models are, independently of liability of major AI proprietors or individual developers and deployers.


This democratization and explosion of information blur the lines between reality and artificial constructs, in echoing Baudrillard's notion of "\textit{hyperreality}". The hyperreality conceived by Baudrillard—an environment where simulacra blur the boundaries between real and artificial, and virtual identities deontologically supersede their real references —becomes an eerily accurate premonition of a possible AI-saturated infosphere \cite{baudrillard1994simulacra}. As AI-generated content swells, we confront the dual challenge of strenghtening our cognitive ecology to preserve our SC, whilst upholding the open-source principles that have traditionally sparked innovation \cite{hutchins2010cognitive}. 
Despite being awash with information, we are precariously perched on the edge of what James Bridle refers to as a "\textit{New Dark Age}," a paradox where information in our current technological ecosystem  obscures knowledge instead of revealing it \cite{bridle2018new}. 
We must navigate this deontological tipping point, resisting unchallenged acceptance of an AI-driven information ecosystem.

The challenge lies in recognizing and navigating this deontological tipping point: this aligns with Floridi's information ethics framework, which underscores the moral implications of creating, managing, and utilizing information. As remarked before, Floridi stresses how that the quality of our infosphere, or the environment in which information is created, shared, and consumed, profoundly impacts our lives and our moral decisions \cite{floridi2010information, floridi2016handbook}. 
To navigate this new complex infosphere, we must engage with a multi-faceted strategy. First, it necessitates moving forward from merely calling AI systems to adhere to ethical guidelines or exhorting to establish a culture of accountability, transparency, and shared responsibility when AI proprietors are able to influence AI agenda and public opinion \cite{mcquillan2022resisting}.
This shift in approach should involve a critical reexamination of why, within our current 
informational ecology, certain narratives are dominant and universally accepted, and who benefits from this status quo. Such societal introspection might prompt a critical reconsideration of the merits of confining the AI debate and our notion of innovation to a single range of solutions.
Furthermore, we argue that, while public online information sources have proven to be fertile ground for the proliferation of AI technologies, today the wealth of SC at stake might be threatened by a range of epistemic risks that we outline using Floridi's taxonomy \cite{floridi2016semantic}:

\begin{itemize}
    \item \textbf{Loss of SC}: This occurs when there is an oversimplification of complex semantic ideas or when an LLM relies on biased or erroneous explanatory models based on incomplete or distorted input data, resulting in flawed argumentation \cite{bender2021dangers, bian2023drop}. In this case, the value of the semantic content is reduced due to the propagation of inaccurate or misleading information, akin to the spread of propaganda, fake news, or "\textit{alternative facts}" \cite{weidinger2022risks}. Protection against this type of risk necessitates rigorous data curation (such as data provenance, and lineage) and model validation protocols to ensure LLMs generate accurate and reliable information.

    \item \textbf{Unproductiveness and Underuse}: When LLMs are used to replicate semantic content without adding value or facilitating a deeper understanding, it can lead to the stagnation of SC. This can happen when users rely too heavily on LLMs for information generation and consumption while neglecting to actively participate in knowledge sharing and debate. Also, at the core, this underuse of SC might stems from the LLMs' architecture, being able to fetch only data that might be available in accessible online repositories, without yet considering the \textit{'long-tail'} of secondary, related contributions, as well as different perspectives, on a given topic. To guard against this risk, it's essential first to inquire over the role of LLMs as epistemic agents, as well as to foster a culture of critical thinking and active engagement in the discourse, preventing the 'mummification' of SC \cite{hutchins2010cognitive}. 
    
    \item \textbf{Misuse}: LLMs, if not properly calibrated or deployed by malicious actors, can generate content that disrespects, misunderstands, or illegitimately appropriates information \cite{bowman2023eight, bian2023drop}. This misuse, or information expropriation, can lead to the loss of SC. Mitigating this risk requires careful design and tracking over their deployment, with due respect for cultural nuances and contexts. In terms of data, this might be possible also leveraging underrepresented communities to not just moderate, but actively participate in data annotation policies, to mitigate potential biases \cite{navigli2023llms}. In terms of models, intellectual property, trademarks, and measure to ensure accountability shall be established to track responsibles within the development and deployment of generative solutions, also enforced by hard laws, such as the forthcoming EU AI Act or the Liability Directive \cite{ai_liability_direct, draftcompromise}.

    \item \textbf{Depreciation}: The value of SC can depreciate over time, particularly when new LLM-generated information floods the infosphere and obscures or distorts earlier knowledge. Future LLMs models, being trained or fine-tuned in such a stagnating environment, might see an increase in diminished returns over their performance. This could happen by being fed data that are either synthetically produced or, even worse, being produced by a shrunken online community of users that lacks incentives to share and engage in knowledge creation and maintenance given the information accessibility of LLMs. Also connected to underuse, the concept of \textit{Model Dementia} has been recently coined \cite{dementia} to signal how future LLMs training datasets might lead to diminished returns in terms of content richness, intended as forgetting underlying data distributions.

\end{itemize}   


Building on this assumption, our collective reliance on language models as repositories of information might entail a shift in our ethical responsibilities, as we transfer the locus of our communal knowledge from the outward sphere of human discourse to the inward representations within these models.
This shift of direction needs also to be put in context, two additional factors play a key role, being inversely proportionals among them: availability of information and attention. 
With the sheer amount of data being produced by LLMs, we might approach new states of information magnitude. This overabundance of information is overshadowing and possibly distorting pre-existing knowledge, causing the depreciation of SC. It's becoming progressively more demanding to discern useful information or valuable knowledge in the face of this onslaught, which in turn undermines the value derived from it. 
In this new era defined by the \textit{Attention Economy} \cite{DavenportB01}, where human attention is a scarce and coveted resource, the pressure on LLMs to be deployed within work or educational tasks, outreach various audiences, and produce engaging content can inadvertently contribute to this range of risks. As these models strive to produce information that appears coherent and well-expounded - such as also sensationalist AI-generated images or news of public personas, sociopolitical facts etc - the focus might shift from providing comprehensive and nuanced insights to offering quick, often shallow pieces of information. This shift could potentially "flatten" the richness of discourse, leading to apparently more engaging, yet less insightful information being circulated.

At the core of this acceleration, the parameter of epistemic filters becomes paramount. These are mechanisms that people use to sort and interpret the information they encounter. They help us decide what counts as evidence for forming a belief or what challenges it enough to lead to belief revision. There are different kinds of filters, among which the two most important ones are filters for omission and discredit \cite{ferrari2020verita}.
Filters for Omission allow individuals or groups to ignore or reject information that does not align with their current beliefs or values. 
Filters for Discredit, instead, lead individuals or groups to dismiss or discredit opposing viewpoints or evidence. This can involve casting doubt on the source of the information, its credibility, or its relevance. Discredit filters are particularly active in polarized debates where individuals or groups have strong beliefs that they feel are being threatened.
Social media platforms, through their algorithmic selection of content, might inadvertently strengthen that. Thus, from there, it can be stated that when encountered with new information, we can actively engage in accepting or rejecting that, yet, a role in such a selection is performed by social aspects of communication.  
In this regard, the concept of \textit{Epistemic Fitness} refers to the effectiveness of an individual's or group's ability to process evidence and revise their beliefs accordingly. It involves the ability to gather, evaluate, and use information to form accurate beliefs about the world. Maximizing epistemic fitness consists in enhancing one's epistemic filters to improve the quality of information intake and the efficiency of belief formation and revision. Through this lens, we can now reevaluate how the AI narratives \cite{lakoff2014all} over capabilities of generative AI are spread by communities with certain interests and beliefs: while some might foresee economic revenues from instilling certain narratives, others might adopt a stance geared toward protecting human rights, remarking the nuances and challenges of human language among others.
From there, we yet have to tackle how to deal with future conversations where LLMs could be deployed to reinforce existing viewpoints, possibly underpinning the deployment of these filters if online users will be led to believe that information spread by LLMs is actually factual and representative of an allegedly major group of people than it is in reality.  


The call to action is thus twofold. On one hand, consumers of AI-generated content need to refine their individual epistemic filters to navigate this new information landscape effectively. This might entail questioning why certain narratives are spread and validated, and for which purposes. On the other hand, developers and proprietors of LLM solutions have an ethical responsibility to design systems that support, rather than undermine, the collective epistemic fitness of society. Deployers, similarly, shall use these tools cognizant of the value of public SC, being also subjected to watermarks, licensing, and any other enforcement to reprove their own accountability.

Thus, to conclude, a cornerstone in our collective response to these risks is the amplification of AI literacy initiatives. Creating an informed citizenry that understands AI technologies, including their potential advantages and associated risks, enables individuals to engage in meaningful discussions and decision-making processes concerning their epistemic validity.
Central to this endeavor is the proactive integration of ethical considerations. Ethical responsibility should not be a reactionary measure or an isolated response to negative outcomes (e.g. regulate only when meaningful harm occurs). Instead, it needs to be woven into the fabric of the AI design and deployment process. Such proactive ethical responsibility can serve as a safeguard, aligning the development and utilization of AI technologies, and disincentivizing diminishing time-to-market agendas.
However, this inquiry does not suggest a departure from open-source practices. Rather, it signals the need for a matured, conscientious version of open-source, devoid of narratives and utopias of technological emancipation or determination. One that is sober, cognizant of the social epistemic risks, and dedicated to enhancing public comprehension of AI technologies.

\section{Conclusion}


This work attempts to evaluate the complex interplay between LLMs' potential for knowledge democratization and the sociotechnical challenges they present. Amid the accelerating proliferation of LLMs in 2023, the widespread narrative that frames them as precursors to AGI risks overshadowing important socio-economic implications, potentially facilitating an AI monopoly. It is vital, therefore, to question who benefits from these narratives and whether these beneficiaries align with societal interests broadly.
Despite acknowledging the lively nature of this debate, we attempt to explore the delicate balance between the democratization of knowledge and the emergence of a deontological tipping point in our infosphere. This tipping point symbolizes a critical juncture where our commitment to open information dissemination may intersect, and potentially conflict, with our obligation to prevent harm. This dynamic has been exacerbated by the cognitive deluge driven by AI technologies, especially LLMs, leading to uncharted social epistemic challenges that stem from their sociotechnical risks.
We have highlighted that the unchecked expansion and proliferation of AI-generated content such as textual information from LLMs, while holding considerable promise, also pose significant risks. Aside from the engaging debate over their properties to handle semantic information (i.e., "understanding"), we shall not fail to commit to a broader inquiry over the ecosystem that fuels attention towards them, being cognizant of a different array of risks that ultimately affect the value of our SC.
\bibliographystyle{unsrt}  
\bibliography{references}

\end{document}